\documentclass[aps,twocolumn,prl,superscriptaddress,showpacs,tightenlines]{revtex4}
\usepackage{amsmath}
\usepackage{graphicx}
\usepackage{epsfig}
\usepackage{subfigure}
\usepackage{amsfonts}

\begin{document}

\title{Birefringence lens effects of atom ensemble
\\ enhanced  by electromagnetically induced transparency}
\author{H. R. Zhang}
\affiliation{Institute of Theoretical Physics, Chinese Academy of
Sciences, Beijing, 100080, China}
\author{Lan Zhou}
\affiliation{Department of Physics, Hunan Normal University,
Changsha 410081, China}
\author{C. P. Sun}
\affiliation{Institute of Theoretical Physics, The Chinese Academy
of Sciences, Beijing, 100080, China}

\begin{abstract}

We study the optic control for birefringence of a polarized light by
an atomic ensemble with a tripod configuration, which is mediated by
the electromagnetically induced transparency with a spatially
inhomogeneous laser. The atom ensemble splits the linearly polarized
light ray into two orthogonally-polarized components, whose
polarizations depend on quantum superposition of the initial states
of the atom ensemble. Accompanied with this splitting, the atom
ensemble behaves as a birefringent lens, which allows one polarized
light ray passing through straightly while focus another orthogonal
to this polarization with finite aberration of focus.
\end{abstract}


\maketitle \narrowtext

\emph{Introduction.} An atom ensemble, manipulated by
electromagnetical field, exhibits various quantum coherent
properties, such as, extremely slow group
velocities~\cite{hau,kash,bud}, large refractive indexes~\cite{mat},
giant nonlinearities~\cite{harris1,Kunch77,xiaomL99}, laser-induced
birefringence~\cite{yoon}, and electromagnetically induced
transparency (EIT)~\cite{harris,arim,gea,xiao,xiaoA71}. Recently, an
enhanced deflection for an unpolarized light beam is observed in an
EIT medium with an external gradient field~\cite{karpa,saut}.
Different from conventional EIT studies, the external fields, used
to control the light propagation, are transversely-inhomogeneous
~\cite{MoseL74,yanA72,karpa,saut,lu,lujing,lan,lan1}.

The EIT-induced light deflection has been explained using quantum
approach with dark state polariton possessing an effective magnetic
moment~\cite{karpa,lan}, or the semiclassical approach~\cite{lu}
with the gradient-index medium, caused by the external fields with
inhomogeneous profiles. The quantum approach in Ref.~\cite{lan}
exhibits the wave-particle duality of the dark polaritons, where an
effective Schr\"{o}dinger equation is derived to describe the
EIT-enhanced spatial motion of the probe field, similar to spinless
particle in an inhomogeneous field.

Most recently, we visualized a polarized light ray as a spin and
study an optical analog of the Stern-Gerlach effect for this ray.
Here, the atom medium becomes anisotropic with respect to the light
polarization when a transversely inhomogeneous field~\cite{lan1} is
applied. Therefore, the linearly polarized probe light splits into
two with opposite circular-polarization. In this case the optical
split can not be simply compared with the original Stern-Gerlach
effect since the incoherent atomic population has been assumed and
thus the split is not a superposition of two polarization exactly.
Actually, such field-induced birefringence have been studied
extensively~\cite{cpol0,cpol1,cpol7,peng74}, but most of them can
refer to the incoherent optical Stern-Gerlach effect.

In this paper, we consider what would happen if the atoms are
initially superposed with submanifold states of the atoms. This
consideration for atomic coherence offers the possibility to change
the polarization of the outgoing wave from linear to any desired
polarization state. It will be showed that, for the atoms with a
tripod configuration, our optically control of polarization is a
direct result using the superpositions of submanifold states with
the intrinsic double-$\Lambda$-type EIT structure. As a laser drives
the atom ensemble to become an anisotropic medium, the EIT assisted
spatial motion of a linearly-polarized probe beam exhibits an
birefringent phenomena with the polarization focusing (defocusing).
Namely, a linearly-polarized light propagates along a straight line
in the medium, another orthogonal-polarized beam comes to a focus at
different positions along the $z$-axis. Such phenomenon -- light
rays parallel to a lens axis fail to converge to the same point, is
called aberration. It is more interesting that, by changing the
frequency of the control light from ``blue detuning'' to ``red
detuning'', the lens-like object can be adjusted optically form
negative (or diverging ) to positive (or converging)cases. This
divergence-to-convergence transition of the lens like effect is only
enhanced in the double EIT configuration.
\begin{figure}[th]
\includegraphics[bb=42 514 543 746,width=8cm]{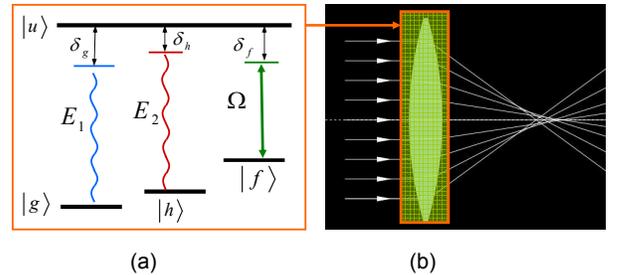}
\caption{(Color online)Energy level scheme (a) for the tripod atoms
interacting with a coupling field (indicated by Rabi frequency
$\Omega$) and a linear-polarized probe field. Such atom ensemble
confined in a gas cell behaves as a converging (or diverging) lens.}
\label{fig:pc1}
\end{figure}

\emph{Setup.} We consider an ensemble of $2N$ identical and
noninteracting
atoms, which is confined in a gas cell of length $L$ along $z$-axis in Fig.~%
\ref{fig:pc1}(b). The atoms possess four levels in a tripod
configuration as shown in Fig.~\ref{fig:pc1}(a). The submanifold of
ground state are spanned by two degenerate Zeeman sublevels
$\left\vert g\right\rangle $ and $\left\vert h\right\rangle $. The
atoms are initially prepared in the superposition $\left\vert \phi
\right\rangle =\alpha \left\vert g\right\rangle +\beta \left\vert
h\right\rangle $ of the Zeeman sublevels. The linearly-polarized
probe light is characterized by field operators $\tilde{E}_{1}$ and
$\tilde{E}_{2}$. Due to some selection rule, the $\sigma
^{+}$-($\sigma ^{-}$-) component $\tilde{E}_{1}$($\tilde{E}_{2}$)
only couples the Zeeman sublevel $\left\vert g\right\rangle
(\left\vert h\right\rangle )$ to excited state $\left\vert
u\right\rangle $, while the transition $\left\vert f\right\rangle
$-$\left\vert u\right\rangle $ is driven by an intense, classical
laser field with a Rabi frequency $\Omega =\Omega (x,y)$, whereas
the transverse component of the intense laser field has a
spatially-inhomogeneous profile. The control beam is detuned from
state $\left\vert u\right\rangle $ with detuning $\delta _{f}$,
while the $\sigma ^{+}$- and $\sigma ^{-}$-polarized components
$\tilde{E}_{1}$ and $\tilde{E}_{2}$ have finite detuning $\delta
_{g}$ and $\delta _{h}$ respectively, which can be adjusted by a
magnetic field applied along the $z$-direction. Obviously, there
exist two $\Lambda $ configurations consisting of energy levels
$(g,f,u)$ and $(h,f,u)$, thus comprises the double-EIT set-up.

Without loss of the generality, we consider the case that both
quantum and classical fields propagate parallel to the $z$-axis. In
reality, both the probe and control field are characterized by wave
packets with spatially-Gaussian profiles. Therefore, each component
of the probe field can be interpreted by a plane wave with a slowly
varying operator~\cite{flei1}
\begin{equation}
\tilde{E}_{j}^{+}(\mathbf{r},t)=\sqrt{\frac{\nu }{2\varepsilon _{0}V}}%
E_{j}\left( \mathbf{r},t\right) e^{i(kz-\nu t)},\left( j=1,2\right) .
\label{pcI-01}
\end{equation}%
We also introduce the following collective continuous operator
$\tilde{\sigma}_{\mu \nu }\left( \mathbf{r}\right) =\sum_{r_{j}\in
N_{r}}\sigma _{\mu \nu }^{j}/N_{r}$ for the collective
excitations in the atomic medium. It actually describes the average of $%
\sigma _{\mu \nu }^{j}=\left\vert \mu \right\rangle _{j}\left\langle \nu
\right\vert $ over $N_{r}$($=\left( 2N/V\right) dr\gg 1$) atoms in a small
but macroscopic volume $V$ around position $\mathbf{r}$. The slowly varying
operators $\sigma _{\mu \nu }$ for the atomic transition operator are
respectively defined as $\tilde{\sigma}_{ug}=\sigma _{ug}\exp (-ikz),\tilde{%
\sigma}_{ug}=\sigma _{ug}\exp (-ikz)$ and
$\tilde{\sigma}_{uf}=\sigma _{uf}\exp (-ik_{c}z)$. Here, $k$ and
$k_{c}$ are the wave numbers to the central frequencies $\nu $ and
$\nu _{c}$ of the probe and control field respectively. For cold
atoms, the kinetic energy could be neglected, so the total system
can be modeled by the interaction Hamiltonian
\begin{eqnarray}
H_{I} &=&\frac{N}{V}\int d^{3}\mathbf{r}\left[ \left( \delta _{g}\sigma
_{gg}+\delta _{h}\sigma _{hh}+\delta _{f}\sigma _{ff}\right) -\right.
\label{pcI-04} \\
&&\left. \left( \Omega \sigma _{uf}+gE_{1}\sigma _{ug}+gE_{2}\sigma
_{uh}+H.c.\right) \right] \text{.}  \notag
\end{eqnarray}%
Due to the symmetry of the states $|g\rangle $ and $|h\rangle $, the
transition matrix element in the above equation is the same $%
g_{ug}=g_{uh}=g=\left\langle u\right\vert d\left\vert g\right\rangle \sqrt{
\nu /\left( 2\varepsilon _{0}V\right) }$ for both circular components,where $%
\left\langle u\right\vert d\left\vert g\right\rangle $ is the dipole matrix
element.

\emph{Effective motion equation of light.} We follow the effective
equation approach in Ref.~\cite{lan1} to study the atomic response
by assuming  the atomic operators as their average in some initial
state, e.g, $\left\vert \phi \right\rangle $. To elucidate the
induced-lens behavior of the atom ensemble, we consider the
Heisenberg-Langevin equations for the atomic and field operators
with the ground-state coherence relaxation rate $\gamma $ and the
decay rate $\Gamma $ of the excited state $\left\vert u\right\rangle
$, which are introduced phenomenologically. Since the intensity of
the quantum probe field is much weaker than that of the control
field, and the number of photons in the signal pulse is much less
than the number of atoms in the sample, we treat the atomic
equations perturbatively with the perturbative parameters
$gE_{i}$~\cite{lan1}. For the initially superposition $\left\vert
\phi (0)\right\rangle $ of Zeeman sublevels, the averages of atomic
operators up to the zeroth order are given by $\sigma
_{mn}^{(0)}=Tr(\left\vert \phi \right\rangle \left\langle \phi
\right\vert \sigma _{mn})$, namely, $\sigma _{gg}^{(0)}=\left\vert
\alpha \right\vert ^{2},\sigma _{hh}^{(0)}=\left\vert \beta
\right\vert ^{2}$ and $\sigma _{gh}^{(0)}=\alpha \beta ^{\ast }$.

With the above consideration, a straightforward calculation with adiabatic
approximation~\cite{lan1} gives the steady-state solution of the atomic
linear response
\begin{subequations}
\label{pcI-09}
\begin{eqnarray}
\sigma _{gu}^{(1)} &=&\zeta _{g}(\left\vert \alpha \right\vert
^{2}E_{1}+\alpha \beta ^{\ast }E_{2}), \\
\sigma _{hu}^{(1)} &=&\zeta _{h}(\left\vert \beta \right\vert
^{2}E_{2}+\alpha ^{\ast }\beta E_{1}).
\end{eqnarray}%
where $\zeta _{s}=g(\delta _{s}-\delta _{f})/[2|\Omega (x,y)|^{2}]$
for $s=g,h$. The above equation shows that when light travels
through an atomic ensemble, the atom response produces collective
electric-dipole moments. This atom response also gives a back-action
on light, thus leads to the paraxial wave equation
\end{subequations}
\begin{equation}
(i\partial _{t}+ic\partial _{z}+\frac{c}{2k}\nabla _{T}^{2})\left[
\begin{array}{c}
E_{1} \\
E_{2}%
\end{array}%
\right] =-2g^{\ast }N\left[
\begin{array}{c}
\sigma _{gu}^{(1)} \\
\sigma _{hu}^{(1)}%
\end{array}%
\right] ,  \label{pcI-10}
\end{equation}%
where $\nabla _{T}^{2}=\partial ^{2}/\partial x^{2}+\partial ^{2}/\partial
y^{2},c$ is the light velocity in vacuum.

Consider the case with degenerate sublevels and no magnetic field,
which means $\delta _{g}=\delta _{h}$. Equations (\ref{pcI-10})
yield a two-component equation
\begin{equation}
\left( i\partial _{t}+ic\partial _{z}+\frac{c}{2k}\nabla _{T}^{2}\right)
\Phi =V(x,y)\sigma ^{(0)}\Phi   \label{pcI-11}
\end{equation}%
for the light field envelope $\Phi =(E_{1},E_{2})^{T}$, which behave as a
spinor moving in a spin-dependent effective potential.
\begin{equation}
V(x,y)\sigma ^{(0)}=-\frac{\left\vert g\right\vert ^{2}N\Delta }{\left\vert
\Omega (x,y)\right\vert ^{2}}\sigma ^{(0)}
\end{equation}%
Here $\Delta =\delta _{h}-\delta _{f}$ is the two photon detuning.
This visualized ``spin'' represents the polarization state of a
probe light. Obviously, the coupling between atoms and light induces
a spin-dependent potential $V(x,y)\sigma ^{(0)}$ to affect light
propagation with opposite polarized-orientation. Consequently, a
signal pulse parallel to the control beam may deviate from its
original trajectory, when it travels across the medium. We also note
that by applying position-dependent fields, an initially isotropic
medium becomes anisotropic~\cite{lan1}. However, the magnetic field
is necessary for the system to display the circular birefringence.
Here, we show that a linear birefringence may also occur.

To describe the propagation of the probe beam clearly, we introduce two
polarized components $E_{-}\equiv \beta E_{1}-\alpha E_{2}$ and $E_{+}\equiv
\alpha E_{1}+\beta E_{2}$ ,which are the coherent superpositions of  the
left- and right-circular polarizations. In the following we only consider
that case  $\alpha $ and $\beta $ are real. In terms of $E_{\pm }$, the Schr%
\"{o}dinger-like equation~(\ref{pcI-11}) becomes
\begin{equation}
\left( i\partial _{t}+ic\partial _{z}+\frac{c}{2k}\nabla _{T}^{2}\right)
E_{\pm }=\frac{1\pm 1}{2}V(x,y)  \label{pcI-15}
\end{equation}%
The above equation indicates that, the $E_{-}$ ray propagates along a
straight line (since no potential act on it ), which means that the medium
is homogenous for $E_{-}$. However, the component $E_{+}$, subject to an
effective potential, experiences a declination.

In order to elucidate the spin-dependent lens behavior induced by the
coupling laser, we assume that the coupling field has a Gaussian profile $%
\Omega \left( r\right) =\Omega _{0}\exp (-x^{2}/2\sigma ^{2})$. Let the
probe beam possess a Gaussian profile
\begin{equation}
E_{\pm }\left( 0\right) =\frac{1}{\sqrt{\pi b^{2}}}e^{-\frac{\left(
x-a\right) ^{2}}{2b^{2}}-\frac{z^{2}}{2b^{2}}}  \label{pcI-17}
\end{equation}%
before it encounters the medium. Here, $\sigma $ is the width of the
driving-field profile, $b$ is the width of the probe field, and $a$ is the
initial location of the wave packet center of the probe field along the $x$
direction. When $b$ is much smaller than $\sigma $, $\left\vert \Omega
\right\vert ^{-2}$ is expanded in Taylor series around $a$, and we retain up
to the linear term. Equation~(\ref{pcI-15}) reads
\begin{eqnarray}
i\partial _{t}E_{\pm } &=&-\left( ic\partial _{z}+\frac{c}{2k}\partial
_{x}^{2}\right) E_{\pm }  \notag  \label{pcI-18} \\
&&-\frac{1\pm 1}{2}\zeta (x-a+\eta /\zeta )E_{+}
\end{eqnarray}%
where $\zeta =2a\eta /\sigma ^{2}$ and $\eta =\Delta \left\vert
g\right\vert ^{2}N\exp (a^{2}/\sigma ^{2})/\Omega _{0}^{2}$. Here,
we restrict our discussion to the 2D system in a $x-z$ plane.
Equation~(\ref{pcI-18}) can be exactly solved by the Wei-Norman
algebraic method~\cite{noma} to give $E_{-}\left( x,t\right)
=E[t,a]$ and
\begin{equation}
\label{pcI-20} E_{+}\left( x,t\right) =E(t,a+\frac{t^{2}\zeta
}{2m})e^{it[\eta -\frac{t}{3}\frac{\zeta ^{2}}{2m}+\zeta (x-a)]}
\end{equation}
where
\begin{equation*}
E[t,a]=\frac{\exp \left[ -\frac{\left( x-a\right)
^{2}}{2(b^{2}+\frac{it}{m}) }-\frac{\left( z-ct\right)
^{2}}{2b^{2}}\right] }{\sqrt{\pi \left( b^{2}+\frac{it}{m}\right)}}
\end{equation*}%
and we have defined the effective mass $m=k/c$.

\emph{Polarization dependent deflection and birefringence lens
effects.} Now we discuss the physics implied in Eq.~(\ref{pcI-20}).
Equation~(\ref{pcI-20}) shows that, the profile center $\left(
x,z\right) =\left( a,0\right)$ of the linear-polarized component
$E_{+}$ at time $t=0$ , is shifted to $(x=x_{+},z=L)$ at time
$t=L/c$, where
\begin{equation}
x_{+}=a+\frac{g^{2}Na\Delta L^{2}}{ck\Omega _{0}^{2}\sigma ^{2}}e^{\frac{%
a^{2}}{\sigma ^{2}}},
\end{equation}
but nothing happens to the $E_{-}$-component. Therefore, if one
tracks the center motion of the probe beam, the trajectory of the
linear-polarized $E_{-}$-component is only a straight line along
$z$-axis. However, its orthogonal-component is deflected by the atom
ensemble. $E_{+}$ propagates either toward or away from the
$z$-axis, which depends on the sign of the two-photon detuning
$\Delta $ and the incident position $a$ of the probe field.
\begin{figure}[th]
\includegraphics[bb=11 616  587 824,width=8 cm]{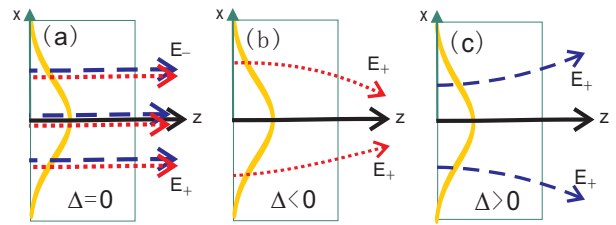}
\caption{(Color online) Schematic of a probe ray passing through a
EIT medium. The vertical line (black one) is the $z$-axis, the
solid-yellow line indicates the profile of the coupling field, the
green rectangle represents the atomic medium. The propagation of the
$E_{-}$-component is illustrated by the blue-dashed line in (a). The
red-dotted line in (a) indicates the trajectory of the
$E_{+}$-component when $\Delta =0$ or $a=0$. The medium
render a lens-like effect in (b) and (c). The medium converges the $E_{+}$%
-component beam (the red-dotted line in (b)) when $\Delta <0$, and diverges
the $E_{+}$-component beam (the blue-dashed line in (c)) when $\Delta >0$. }
\label{fig:pc2}
\end{figure}

Figure~\ref{fig:pc2} shows the different cases for the $E_{\pm
}$-ray propagation. Panel~\ref{fig:pc2}(a) indicates the trajectory
of the linear-polarized component $E_{-}$ (blue-dashed line) as well
as its orthogonal component $E_{+}$ (red-dotted line) under the
situation $\Delta =0 $ or $a=0$. They all travels across the atomic
medium straightly. Panel~\ref{fig:pc2}(b) corresponds to the case
when $\Delta <0$, where the medium acts as a lens causing focus of
the $E_{+}$-component. Panel~\ref{fig:pc2}(c) describes the case
when $\Delta >0$, where the $E_{+}$-component experiences a
defocusing.

\begin{figure}[th]
\includegraphics[bb=68 487 523 734,width=7 cm]{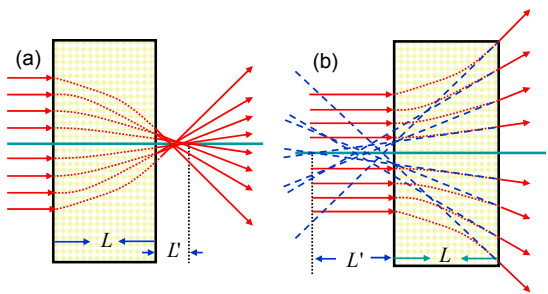}
\caption{(Color online) Schematic of an optic-controlled lens and its
aberration. (a) The EIT medium converges all input parallel rays to
different points along the $z$-axis. (b) The EIT medium diverges all input
parallel rays, which are brought to different points along the $z$-axis at
the same side of the incident rays.}
\label{fig:pc3}
\end{figure}

The above analysis implies that the EIT medium behaves as a lens with a
varying focus, which can be feasibly optical-controlled. Right after light
leaves the medium, it obtains a transverse group velocity with magnitude
\begin{equation}
v_{x}=\frac{2La\Delta |g|^{2}N}{k\sigma ^{2}\Omega _{0}^{2}}\exp
(a^{2}/\sigma ^{2})
\end{equation}%
In the case of \textquotedblleft red detuning\textquotedblright\ $\Delta <0$
[Fig.~\ref{fig:pc3}(a)], the EIT medium made $v_{x}$ toward the $z$-axis.
Therefore, a collimated light ray, starting at points $x=\pm a$ parallel to
the lens axis (i.e. $z$-axis), will meet the $z$-axis at
\begin{equation}
z=f_{con}\left( a\right) =\frac{L}{2}+F(a).
\end{equation}%
where $F(a)=kc\sigma ^{2}\Omega _{0}^{2}/(2L|\Delta| |g|^{2}N)\exp
(-a^{2}/\sigma ^{2})$. However, as the
distance $a$ changes from $0$ to $h$ (the hight of the cuboid gas
cell), this focal point runs from $f_{con}(h)$ to $f_{con}(0)$. This
shows a typical aberration with a distortion length $l=F(0)-F(h)$.
Obviously, such an EIT-based lens do not form perfect images, due to
distortions or aberrations introduced by the $a$-dependent focal
point and the widths spreading of the wave packet.  Therefore, the
atomic medium functions as a converging (or positive) lens to some
extent. In the "blue detuning " case with ($\Delta >0$ in
Fig.3.(b)), although the transverse velocity has the same magnitude
$v_{x}$, its direction is opposite to the former case. Here, the
light ray will not meet the $z$-axis in the propagating direction,
but it virtually crosses the $z$-axis at $z=z_{div}=-[F(a)-L/2]$,
which means that the atomic medium acts as a diverging lens. Thus
the atom ensemble behaves as a negative or diverging lens, i.e.,
collimated light rays, passing through the medium, is diverged.

Therefore, changing the frequency of the control light from ``blue
detuning'' to ``red detuning'', we can carry out an
optically-controlled quantum manipulation based on EIT for the
transition from the diverging lens effect to converging one. The
light ray experiences either defocusing or focusing determined by
the sign of the two-photon detuning $\Delta $. To realize an
optically-controlled lens of divergence-to-convergence transition,
we would like to consider some experimental data: $\nu =3\times
10^{15}rad/s$, $N/V=10^{13}cm^{-3}$, $\Omega _{0}=0.6\Gamma $,
$L=10cm$, $\sigma =L/4$. For a ray starting at $a=L/4$, the
deflection angle $\alpha \simeq 1.9\times 10^{-2}$ when two photon
detuning $\Delta =0.1\Gamma $. This EIT-gas lens seems different to
put into service currently, but the further improvements with EIT
enhancement \cite{karpa} can lead to the obviously-observable
effect.

\emph{Conclusion.} In this paper, we study the defocusing and
focusing of probe light by an ensemble of four-level atoms in a
tripod configuration driving by a control field with spatially
inhomogeneous profile. We find the atomic medium serves as a
polarization-selective lens for the probe beam. Different from
previous proposal~\cite{lan1}, the Zeeman splitting of magnetic
sub-levels does not cause asymmetry atomic susceptibility for left-
and right-circular polarization components of the optical field
because no magnetic field is applied in our setup. The present
investigation shows that the probe beam still splits into two, and
each polarized component of the outgoing probe field contains the
information of the atomic population. Therefore, preparation for the
atomic internal state can also be used to control the deflection,
focusing and defocusing effects.

This work was supported by the NSFC with Grant No.~90203018,
No.~10775048,No.~10325523, and No.~10704023, and NFRPC with Grant
No.~2006CB921206,and No.~2007CB925204

\end{document}